\documentclass[conference]{IEEEtran}
\IEEEoverridecommandlockouts
\usepackage{cite}
\usepackage{amsmath,amssymb,amsfonts}
\usepackage{algorithmic}
\usepackage{graphicx}
\usepackage{textcomp}

\usepackage{booktabs}
\usepackage{enumitem}
\usepackage{stfloats}

\usepackage{xcolor}
\def\BibTeX{{\rm B\kern-.05em{\sc i\kern-.025em b}\kern-.08em
    T\kern-.1667em\lower.7ex\hbox{E}\kern-.125emX}}

\usepackage{bm}
    
\begin{document}

\def\x{{\mathbf x}}
\def\L{{\cal L}}
\def\mathbi#1{\textbf{\em #1}}

\def\UrlBreaks{\do\/\do-}
\def\mathbi#1{\textbf{\em #1}}

\title{Onset and offset weighted loss function for sound event detection\\
}

\author{
    \IEEEauthorblockN{Tao Song}
    % \IEEEauthorblockA{\textit{Huawei Technologies Co., Ltd.}} \\
    %\textit{Audio engineering department}\\
    % Beijing, China \\
}

\maketitle

\begin{abstract}
In a typical sound event detection (SED) system, the existence of a sound event is detected at a frame level, and consecutive frames with the same event detected are combined as one sound event.
The median filter is applied as a post-processing step to remove detection errors as much as possible. However, detection errors occurring around the onset and offset of a sound event are beyond the capacity of the median filter.
To address this issue, an onset and offset weighted binary cross-entropy (OWBCE) loss function is proposed in this paper, which trains the DNN model to be more robust on frames around onsets and offsets.
Experiments are carried out in the context of DCASE 2022 task 4. Results show that OWBCE outperforms BCE when different models are considered. For a basic CRNN, relative improvements of $6.43\%$ in event-F1, $1.96\%$ in PSDS1, and $2.43\%$ in PSDS2 can be achieved by OWBCE.
\end{abstract}


\section{Introduction}
Sound event detection (SED) is an important research topic in computational auditory scene analysis (CASA), which can be used in various areas\cite{Takahashi_2018, Anastasios_2020, Manikanta_2019, Mnasri_2022}.
Given an audio clip, SED is expected to detect the existence of sound events and estimate corresponding timestamps (onsets and offsets). These two subtasks are usually considered as one single \textit{multi-class multi-label classification} problem and solved with the deep neural network (DNN)\cite{Mesaros_2021}. A general-purpose network architecture for SED is the convolutional recurrent neural network (CRNN)\cite{Cakir_2017, Ebbers_2021, Hyeonuk_2022},
in which CNN extracts high-level features and RNN learns temporal dependencies in the sequence of high-level features.

In a SED system, the input audio clip is first divided into frames, for example, with a frame length of 64 ms. Features (e.g., mel-spectrum) extracted from all frames are fed into the DNN model to detect whether sound events of interest are present in each frame. Finally, consecutive frames with the same sound event detected are combined as one sound event, and the first and last frames correspond to the onset and offset of this sound event.

\begin{figure}[tbp]
  \centering
  \includegraphics[width=\linewidth]{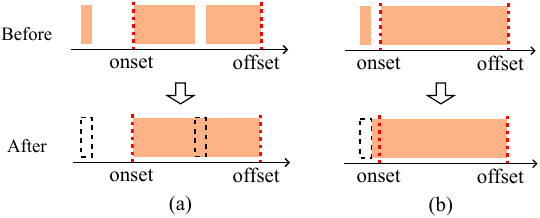}
\caption{Two examples of median filtering on detection results. Detection results are denoted with filled rectangles, and the real onset and offset are denoted with red dashed lines. In (a), detection errors that occur within or away from the sound event are removed by the median filter. In (b), detection errors that occur around the onset are also removed by the median filter, but new errors are introduced. }
\label{fig:median_filter}
\end{figure}

Under the influence of reverberation, overlap between sound events, and other factors, it is challenging to detect whether sound events are present in a frame. Detection results given by the DNN model are likely to contain errors, and a few such errors can lead to a fake sound event or split of a sound event.
For example, if a few frames in the middle of a sound event \textit{Speech} are detected as non-speech, there will be two separated \textit{Speech} rather than one in the final result.
For robustness, the median filter is usually adopted as a post-processing step\cite{Ebbers_2021, Hyeonuk_2022, Park_2022}. Based on the assumption that the duration of a sound event or the interval between sound events must exceed certain values, the median filter can remove detection errors by removing sound events or filling intervals that span only a few frames (see Fig. \ref{fig:median_filter}(a)).
However, there are detection errors that the median filter can not deal with properly. As shown in Fig. \ref{fig:median_filter}(b), if detection errors occur around the onset or offset of a sound event, the median filter can remove existing detection errors but at the cost of introducing new detection errors.

As a post-processing step, the median filter has a limited ability to remove detection errors that occur around the onset and offset of a sound event.
For better performance, especially in timestamp estimation, the DNN model is required to be more robust on frames around the onset and offset of a sound event, which can be achieved by optimizing the DNN model primarily for these frames. Based on this idea, an onset and offset weighted binary cross-entropy (OWBCE) loss function is proposed in this paper.
To our knowledge, such an issue has not been considered before.

The rest of this paper is organized as follows. In section 2, works on weighted loss function in SED are briefly reviewed. The proposed OWBCE is described in section 3. Section 4 presents the experimental setup and evaluation results. Section 5 concludes this paper.

\begin{figure*}[htbp]
	 \centering
	 \includegraphics[width=0.75\linewidth]{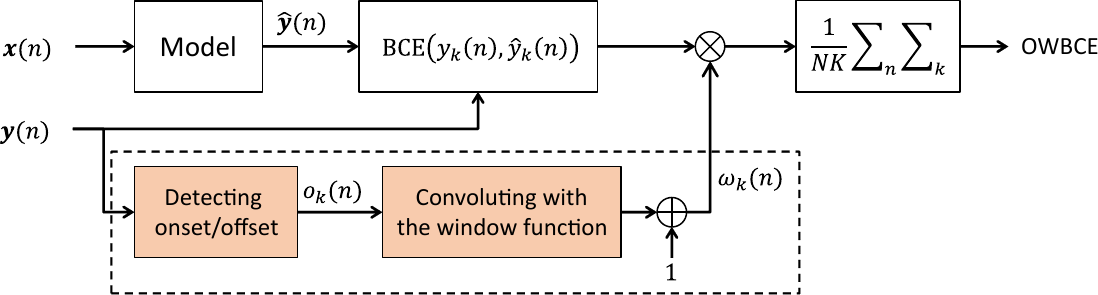}
	\caption{Calculation of OWBCE. }
	% \caption{Calculation of OWBCE. $\mathbi{x}$ and $\mathbi{y}$ are the spectrum and corresponding label. }
	\label{fig:OWBCE}
\end{figure*}

\section{related work}
\label{sec:related_work}
 Assuming there are $N$ audio frames and $K$ event classes, the DNN model is expected to output $N\times K$ estimations. The training loss is calculated for each estimation and then averaged over frames and event classes,
\begin{equation}
	\label{eq:loss_func}
	\mathcal{L}(\mathbi{y}, \hat{\mathbi{y}}) = \frac{1}{NK}\sum_{n}\sum_{k}f\bigl(y_{k}(n), \hat{y}_{k}(n)\bigr)
\end{equation}
where $y_{k}(n)$ and $\hat{y}_k(n)$ are the ground truth and estimation of whether sound event $k$ is present in frame $n$, $f$ is the loss function which can be binary cross-entropy (BCE).
Some works introduce weighting functions into Eq.\ref{eq:loss_func} and proposed weighted loss functions\cite{kiyokawa_2019, Imoto_2021, Park_2022}. These works focus on the data imbalance problem and only consider weighting training loss along the sound event dimension.

Data imbalance is a major problem in SED. Taking Domestic Environment Sound Event Detection (DESED) dataset\cite{Serizel_2020} as an example, in the training subset, the number of sound events \textit{Speech} can be 15 times as large as the number of other sound events (e.g., \textit{Blender}). Such imbalance between sound events can result in a natural bias of trained models towards overrepresented sound events. To address this issue, a loss weighting function is proposed in \cite{kiyokawa_2019}, which is defined as
\begin{equation}
\omega(k) = \frac{e^{1/M_{k}}}{\sum_{k=1}^{K}e^{1/M_{k}}}
\end{equation}
where $M_k$ is the total number of sound events $k$. Inversely proportional to the number of sound events, the weighting function can counterbalance the bias caused by data imbalance to a certain extent.

Apart from the number of occurrences, the duration is another dimension of data imbalance. Since the DNN model is trained at a frame-level, the data volume of each sound event is decided by not only its number but also its duration. To address this compounded imbalance issue, another loss weighting function is proposed in \cite{Park_2022}, which is defined as
\begin{equation}
	\begin{aligned}
	\omega(k) &\propto \frac{1-\beta(k)}{1-\beta(k)^{ \lfloor \lambda \times r(k) \rfloor}}, \quad \sum_{k}\omega(k) = K \\
	\beta(k) &= \frac{M_k-1}{M_k} \\
	r(k) &= \frac{N_k}{\sum_{k} N_k}
	\end{aligned}
\end{equation}
where, $\lambda$ is a hyper-parameter, $N_k$ is the total frame number of sound event $k$.  Similar weighting functions are also proposed in \cite{Imoto_2021}.

\section{Proposed method}
\label{sec:method}

To train a DNN model that is more robust on frames around onsets and offsets, the onset and offset weighted binary cross-entropy(OWBCE) loss function is proposed in this paper.
Different to existing works, the training loss is weighted along the frame dimension, and the weighting function is calculated individually for each input audio clip.

\begin{figure}[htp]
  \centering
  \includegraphics[width=0.7\linewidth]{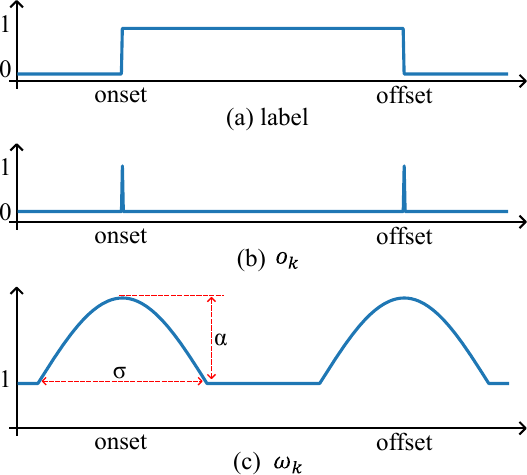}
\caption{A example of the loss weighting function. }
\label{fig:OWBCE_eg}
\end{figure}

The calculation of OWBCE is illustrated in Fig. \ref{fig:OWBCE}.
In the upper path, BCE loss is calculated for each frame and each class. The weighting function is calculated from labels in two steps in the lower path.
In the first step, onsets and offsets in labels are located using first-order difference. As shown in Fig. \ref{fig:OWBCE_eg}(b), each onset or offset is indicated by a Dirac delta function. 
If the input audio clip is annotated with soft labels, the amplitude of labels can be preserved in the Dirac delta function.
For event class $k$, the result of onset and offset detection is denoted as $o_k$.

In the second step, $o_k$ is convoluted with a \textit{sin} window and increased by 1, as shown in Fig. \ref{fig:OWBCE_eg}(c). Since the convolution results outside onset or offset regions equals 0, the weighting function is further increased by 1 to ensure that training loss in these regions is not affected
The height and width of the \textit{sin} window, denoted as $\alpha$ and $\sigma$, are two hyper-parameters. $\alpha$ determines how much weight will be applied to frames around the onset and offset, while $\sigma$ determines how many frames around the onset and offset will be weighted. The effect of these two parameters will be explored in the following experiments.

\section{Experiment}
\label{sec:experiment}

The performance of OWBCE is examined in three experiments. Two hyper-parameters, $\alpha$ and $\sigma$, are optimized in the first experiment. The second experiment examines the robustness of OWBCE to human errors in timestamp annotations. The first two experiments are carried out using a basic CRNN model. In the final experiment, the effectiveness of OWBCE is validated using two other models.

\subsection{Dataset}

All three experiments are carried out based on the DESED dataset, which is composed of 10 sound event classes. The training subset contains 10,000 strong-labeled audio clips, 1578 weak-labeled audio clips, and 14,412 unlabeled audio clips. The weak-labeled and unlabeled audio clips are recorded in domestic environments. %taken from Audioset\cite{Jort_2017}.
The strong-labeled audio clips are synthesized by mixing individual sound event instances with background audio.  All assessments are performed on the validation subset, which consists of strong-labeled recordings.

In addition to DESED, the baseline of DCASE 2022 task 4 provides 3470 strong-labeled recordings, which will be used in the second experiment.

\subsection{Experimental setting}
The log-mel spectrogram is used as the input feature of the DNN model, which is extracted on 16 kHz audio with 128 frequency bins, 2048 window length, and 256 hop length. As a result, the spectrogram has a time resolution of 16 ms.

In the training phase, the same data augmentation scheme is adopted as in \cite{Hyeonuk_2022}, which consists of frame shift, Mixup\cite{Hongyi_2018}, time masking, and FilterAugment\cite{Nam_2022}. For each batch of training data, the training loss is a weighted sum of three parts,
\begin{equation}
	\mathcal{L} = \mathcal{L}_{strong} + \omega_{weak}\mathcal{L}_{weak} + \omega_{cons}\mathcal{L}_{cons}
\end{equation}
where, $\mathcal{L}_{strong}$ and $\mathcal{L}_{weak}$ are BCE loss calculated on strong-labeled and weak-labeled data, $\mathcal{L}_{cons}$ is consistency loss for mean-teachers, $\omega_{weak}$ and  $\omega_{cons}$ are two hyper-parameters. The proposed loss weighting function is only applied to $\mathcal{L}_{strong}$. % A warm-up strategy is adopted.
In the inferring phase, the output of the DNN model is first binaried and then median filtered. For each event class, a decision threshold of 0.5 and event-specific median filter length are used.

A basic CRNN is used in the first two experiments. This CRNN is the same model used in the baseline of DCASE 2022 task 4, except that the kernel number of each CNN layer is doubled, and ReLU is replaced with Context Gating. The proposed loss function is tested in the third experiment using the latest models, FDY-CRNN\cite{Hyeonuk_2022} and SK-CRNN\cite{He_2022}.

For each condition, an average performance of 5 runs is reported. Three performance metrics from DCASE 2022 task 4 are adopted, which are event-based f-score (event-F1) and polyphonic sound event detection score (PSDS)\cite{Bilen_2020} with two different settings (PSDS1 and PSDS2). Event-F1 is a boundary-based metric, while PSDS is an intersection-based metric. Details on these metrics can be found in \cite{DCASE2022_task4}.

\subsection{Hyper-parameter optimization}
In this experiment, hyper-parameters, $\alpha$, and $\sigma$ are optimized by training and evaluating models for each possible combination of values.

To narrow down the search space, these two hyper-parameters are optimized independently and based on preliminary experiment results. 

\begin{figure}[htbp]
  \centering
  \includegraphics[width=\linewidth]{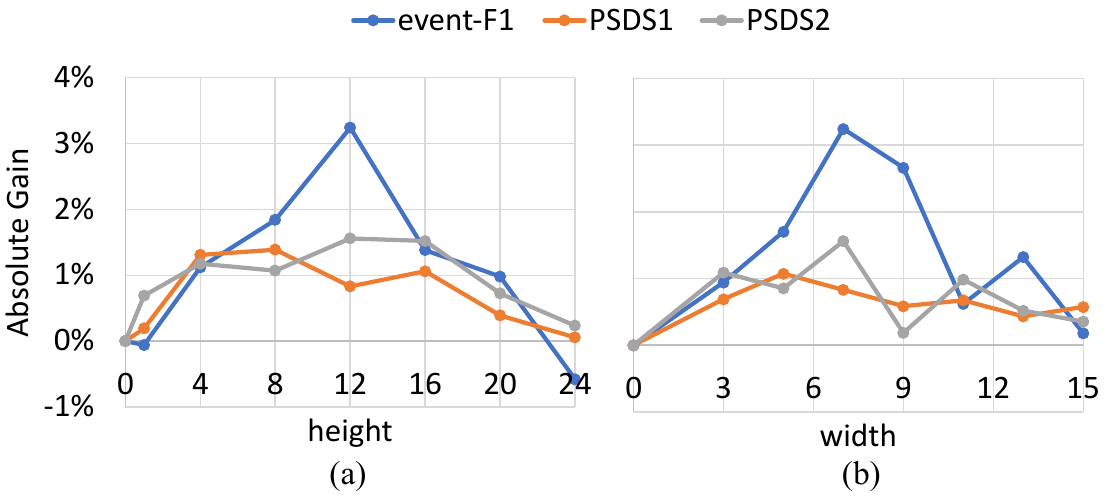}
\caption{Performance gain of OWBCE to BCE as a function as the width and height of \textit{sin} window. Width is set to 7 in (a) and height is set to 12 in (b).}
\label{fig:performance_exp1}
\end{figure}

In the first step, $\alpha$ is optimized with $\sigma$ temporally set to 7. The performance gain of OWBCE to BCE as a function of $\alpha$ is plotted in figure \ref{fig:performance_exp1}(a). It should be noted that OWBCE with  $\sigma=0$ or $\alpha=0$ is equivalent to BCE. On average, the best performance is reached when $\alpha=12$.

In the second step, $\sigma$ is optimized with $\alpha$ set to 12. The performance gain as a function of $\sigma$ is plotted in figure \ref{fig:performance_exp1}(b). On average, the best performance is reached when $\sigma=7$. The input spectrogram has a time resolution of 16 ms and is down-sampled by a factor of 4 along the time dimension in CRNN. As a result, the detection result has a time resolution of 64 ms. $\sigma=7$ means the half of the window width is 192 ms, which is approximately equal to the collar (200 ms) used in event-F1.
In sum, the optimum width and height of \textit{sin} window are $7$ and $12$, respectively.

The final performances of OWBCE and BCE are listed in Table \ref{tab:performance_exp1}. OWBCE outperforms BCE in all three metrics. Compared to BCE, OWBCE improves event-F1 by $6.43\%$, PSDS1 by $1.96\%$ and PSDS2 by $2.43\%$. The improvement of event-F1 is more than twice as large as the improvements of the other two metrics. Such results show that OWBCE can improve the accuracy of timestamp estimations.

\renewcommand\arraystretch{1.2} 
\begin{table}[htbp]
\centering
\caption{Performance of OWBCE and BCE on DESED dataset. The DNN model used is a basic CRNN.}
\label{tab:performance_exp1}
\begin{tabular*}{\hsize}{@{\extracolsep{\fill}}cccc@{}}
\toprule
     & event-F1(\%) & PSDS1(\%) & PSDS2(\%) \\ \midrule
%BCE  & $50.40\pm0.47$        & $42.28 \pm 0.71$     & $64.26 \pm 0.28$    \\
%OWBCE & $\mathbf{53.64 \pm 0.60}$        & $\mathbf{43.11 \pm 0.40}$     & $\mathbf{65.82 \pm 0.24}$     \\ \midrule
BCE  & $50.40$        & $42.28$     & $64.26$    \\
OWBCE & $\mathbf{53.64}$        & $\mathbf{43.11}$     & $\mathbf{65.82}$     \\ \midrule
% gain & 3.24         & 0.83      & 1.56      \\ \bottomrule
\end{tabular*}
\end{table}

\subsection{Robustness to human errors in timestamp annotation}

The performance of OWBCE relies on the accuracy of onset and offset annotations. 
Compared to synthesized data, manually labeled timestamps are likely to contain errors, given the ambiguity in the perception of onsets and offsets. To explore the robustness of OWBCE to human errors, the synthesized data in DESED training subset is replaced with manually labeled real recordings provided by the baseline of DCASE 2022 task 4.

The performance of OWBCE and BCE on the modified DESED dataset is shown in Table \ref{tab:performance_exp2}.
Compared to BCE, OWBCE improves event-F1 by $4.75\%$, PSDS1 by $2.55\%$ and PSDS2 by $2.01\%$. OWBCE outperforms BCE in all three metrics, which proves its robustness.

\renewcommand\arraystretch{1.2} 
\begin{table}[htbp]
\centering
\caption{Performance of OWBCE and BCE on modified DESED data. The DNN model used is a basic CRNN.}
\label{tab:performance_exp2}
\begin{tabular*}{\hsize}{@{\extracolsep{\fill}}cccc@{}}
\toprule
     & event-F1(\%) & PSDS1 (\%) & PSDS2(\%) \\ \midrule
BCE  & 53.06        & 42.69     & 65.28     \\
OWBCE & \textbf{55.58}      & \textbf{43.78}     & \textbf{66.59}     \\ \bottomrule
\end{tabular*}
\end{table}

Comparing Table \ref{tab:performance_exp2} to Table \ref{tab:performance_exp1}, it can be found that models trained with strong-labeled recordings achieve better performance than models trained with synthesized data; this may be because synthesized data fails to capture the complexity of real-life environments.

\subsection{Effectiveness on other models}
The benefit of OWBCE has been proved in the previous two experiments, but the result is limited to a basic CRNN model. In this experiment, two of the latest models are considered,  which share the same architecture as CRNN but use specially designed CNN layers.

In SK-CRNN, the selective kernel (SK) unit is used, which can adjust receptive field size based on multiple scales of its input. The performance of OWBCE and BCE is listed in Table \ref{tab:performance_exp3_SKN}. OWBCE outperforms BCE in event-F1, but the performance in PSDS1 and PSDS2 are comparable. The relative improvement of event-F1 is $3.66\%$.

\renewcommand\arraystretch{1.2} 
\begin{table}[htbp]
\centering
\caption{Performance of OWBCE and BCE on DESED dataset, the DNN model used is SK-CRNN.}
\label{tab:performance_exp3_SKN}
\begin{tabular*}{\hsize}{@{\extracolsep{\fill}}cccc@{}}
\toprule
     & event-F1(\%) & PSDS1(\%) & PSDS2(\%) \\ \midrule
BCE  & 50.34        & \textbf{43.56}     & 67.00     \\
OWBCE & \textbf{51.88}        & 43.38     & \textbf{67.12}     \\ \midrule
% gain & 3.24         & 0.83      & 1.56      \\ \bottomrule
\end{tabular*}
\end{table}

In FDY-CRNN, frequency dynamic convolution is adopted, which can recognize frequency-dependent patterns of sound events. The performance of OWBCE and BCE is listed in Table \ref{tab:performance_exp3_FDY}. OWBCE outperforms BCE in event-F1 and PSDS1, but the performance in PSDS2 is a little worse. The relative improvements of event-F1 and PSDS1 are $3.66\%$ and $1.73\%$, respectively. It should be noted that the performance of FDY-CRNN reported in \cite{Hyeonuk_2022} is a little higher. The main reason is that \cite{Hyeonuk_2022} only reports the best performance among 16 runs, while the average performance of 5 runs is reported here. The best performance of FDY-CRNN among 5 runs is comparable to the result reported in \cite{Hyeonuk_2022}.

\renewcommand\arraystretch{1.2} 
\begin{table}[htbp]
\centering
\caption{Performance of OWBCE and BCE on DESED dataset, the DNN model used is FDY-CRNN.}
\label{tab:performance_exp3_FDY}
\begin{tabular*}{\hsize}{@{\extracolsep{\fill}}cccc@{}}
\toprule
     & event-F1(\%) & PSDS1(\%) & PSDS2(\%) \\ \midrule
BCE  & 52.42        & 43.50     & \textbf{65.04}     \\
OWBCE & \textbf{54.34}        & \textbf{44.26}     & 64.11     \\ \midrule
% gain & 3.24         & 0.83      & 1.56      \\ \bottomrule
\end{tabular*}
\end{table}

%In sum, OWBCE outperforms BCE in event-F1 and performances in PSDS1 and PSDS2 are comparable.
In sum, compared to BCE, OWBCE has a better performance in event-F1 and comparable performance in PSDS1 and PSDS2.

\section{conclusion}
An onsets and offsets weighted binary cross-entropy is proposed in this paper, which is denoted as OWBCE.
In a SED system, the input audio clip is divided into frames and classified by a DNN model. The output of the DNN model is median-filtered to remove possible errors. However, detection errors that occur around the onset and offset of sound events are beyond the capacity of a median filter. For better performance, the DNN model should be more robust to frames around onsets and offsets. To address this issue, OWBCE is proposed to optimize the DNN model primarily for frames around onsets and offsets. Three experiments are carried out. The first experiment shows OWBCE outperforms BCE when a basic CRNN model is considered. The second experiment shows that OWBCE is robust to human errors in onset and offset annotations. The third experiment validates the effectiveness of OWBCE on two of the latest CRNN models.

\bibliographystyle{IEEEtran}
\bibliography{OWBCE_SED}

\begin{thebibliography}{10}
\providecommand{\url}[1]{#1}
\csname url@samestyle\endcsname
\providecommand{\newblock}{\relax}
\providecommand{\bibinfo}[2]{#2}
\providecommand{\BIBentrySTDinterwordspacing}{\spaceskip=0pt\relax}
\providecommand{\BIBentryALTinterwordstretchfactor}{4}
\providecommand{\BIBentryALTinterwordspacing}{\spaceskip=\fontdimen2\font plus
\BIBentryALTinterwordstretchfactor\fontdimen3\font minus
  \fontdimen4\font\relax}
\providecommand{\BIBforeignlanguage}[2]{{%
\expandafter\ifx\csname l@#1\endcsname\relax
\typeout{** WARNING: IEEEtran.bst: No hyphenation pattern has been}%
\typeout{** loaded for the language `#1'. Using the pattern for}%
\typeout{** the default language instead.}%
\else
\language=\csname l@#1\endcsname
\fi
#2}}
\providecommand{\BIBdecl}{\relax}
\BIBdecl

\bibitem{Takahashi_2018}
N.~Takahashi, M.~Gygli, and L.~Van~Gool, ``Aenet: Learning deep audio features
  for video analysis,'' \emph{IEEE Transactions on Multimedia}, vol.~20, no.~3,
  pp. 513--524, 2018.

\bibitem{Anastasios_2020}
A.~Vafeiadis, K.~Votis, D.~Giakoumis, D.~Tzovaras, L.~Chen, and R.~Hamzaoui,
  ``Audio content analysis for unobtrusive event detection in smart homes,''
  \emph{Engineering Applications of Artificial Intelligence}, vol.~89, p.
  103226, 2020.

\bibitem{Manikanta_2019}
K.~Manikanta, K.~Soman, and M.~S. Manikandan, ``Deep learning based effective
  baby crying recognition method under indoor background sound environments,''
  in \emph{International Conference on Computational Systems and Information
  Technology for Sustainable Solution (CSITSS)}, 2019, pp. 1--6.

\bibitem{Mnasri_2022}
Z.~Mnasri, S.~Rovetta, and F.~Masulli, ``Anomalous sound event detection: A
  survey of machine learning based methods and applications,'' \emph{Multimedia
  Tools and Applications}, vol.~81, no.~4, pp. 5537--5586, 2022.

\bibitem{Mesaros_2021}
A.~Mesaros, T.~Heittola, T.~Virtanen, and M.~D. Plumbley, ``Sound event
  detection: A tutorial,'' \emph{IEEE Signal Processing Magazine}, vol.~38,
  no.~5, pp. 67--83, 2021.

\bibitem{Cakir_2017}
E.~Çakır, G.~Parascandolo, T.~Heittola, H.~Huttunen, and T.~Virtanen,
  ``Convolutional recurrent neural networks for polyphonic sound event
  detection,'' \emph{IEEE/ACM Transactions on Audio, Speech, and Language
  Processing}, vol.~25, no.~6, pp. 1291--1303, 2017.

\bibitem{Ebbers_2021}
J.~Ebbers and R.~Haeb-Umbach, ``Self-trained audio tagging and sound event
  detection in domestic environments,'' in \emph{Proceedings of the 6th
  Detection and Classification of Acoustic Scenes and Events 2021 Workshop
  (DCASE2021)}, 2021, p. 226–230.

\bibitem{Hyeonuk_2022}
H.~Nam, S.~Kim, B.~Ko, and Y.~Park, ``Frequency dynamic convolution:
  Frequency-adaptive pattern recognition for sound event detection,'' in
  \emph{Interspeech}, Incheon, Korea, Sep. 2022, pp. 2763--2767.

\bibitem{Park_2022}
S.~Park and M.~Elhilali, ``Time-balanced focal loss for audio event
  detection,'' in \emph{International Conference on Acoustics, Speech and
  Signal Processing (ICASSP)}, Singapore, Singapore, Apr. 2022, pp. 311--315.

\bibitem{kiyokawa_2019}
Y.~Kiyokawa, S.~Mishima, T.~Toizumi, K.~Sagi, R.~Kondo, and T.~Nomura,
  ``\BIBforeignlanguage{en}{Sound event detection with resnet and self-mask
  module for dcase 2019 task 4},'' Tech. Rep., 2019.

\bibitem{Imoto_2021}
K.~Imoto, S.~Mishima, Y.~Arai, and R.~Kondo, ``Impact of sound duration and
  inactive frames on sound event detection performance,'' in
  \emph{International Conference on Acoustics, Speech and Signal Processing
  (ICASSP)}, Toronto, Canada, May 2021, pp. 860--864.

\bibitem{Serizel_2020}
R.~Serizel, N.~Turpault, A.~Shah, and J.~Salamon, ``{Sound event detection in
  synthetic domestic environments},'' in \emph{{International Conference on
  Acoustics, Speech, and Signal Processing}}, Barcelona, Spain, May 2020.

\bibitem{Hongyi_2018}
H.~Zhang, M.~Ciss{\'{e}}, Y.~N. Dauphin, and D.~Lopez{-}Paz, ``mixup: Beyond
  empirical risk minimization,'' in \emph{International Conference on Learning
  Representations (ICLR)}, 2018.

\bibitem{Nam_2022}
H.~Nam, S.-H. Kim, and Y.-H. Park, ``Filteraugment: An acoustic environmental
  data augmentation method,'' in \emph{International Conference on Acoustics,
  Speech and Signal Processing (ICASSP)}, Singapore, Singapore, Apr. 2022, pp.
  4308--4312.

\bibitem{He_2022}
K.~He, X.~Shu, S.~Jia, and Y.~He, ``\BIBforeignlanguage{en}{Semi-{Supervised}
  {Sound} {Event} {Detection} {System} {For} {Dcase} 2022 {Task} 4},'' Tech.
  Rep., 2022.

\bibitem{Bilen_2020}
{\c{C}}.~Bilen, G.~Ferroni, F.~Tuveri, J.~Azcarreta, and S.~Krstulovi{\'c}, ``A
  framework for the robust evaluation of sound event detection,'' in
  \emph{International Conference on Acoustics, Speech and Signal Processing
  (ICASSP)}, Barcelona, Spain, Apr. 2020, pp. 61--65.

\bibitem{DCASE2022_task4}
``{DCASE} 2022 task 4: Task description,''
  \url{https://dcase.community/challenge2022/task-sound-event-detection-in-domestic-environments}.

\end{thebibliography}

\vspace{12pt}
%\color{red}
%IEEE conference templates contain guidance text for composing and formatting conference papers. Please ensure that all template text is removed from your conference paper prior to submission to the conference. Failure to remove the template text from your paper may result in your paper not being published.

\end{document}